\documentclass[twocolumn]{aastex61} 
\usepackage{graphicx}
\usepackage{amsmath}
\usepackage{amssymb,latexsym,amsopn}
\usepackage{color}

\usepackage{ulem}

\shorttitle{The FRB spectrum}
\shortauthors{Ravi \& Loeb}

\begin{document}

\title{Explaining the statistical properties of Fast Radio Bursts with suppressed low-frequency emission}

\correspondingauthor{Vikram Ravi}
\email{vikram.ravi@cfa.harvard.edu}

\author{Vikram Ravi}
\affil{Harvard-Smithsonian Center for Astrophysics, 60 Garden Street, Cambridge, MA 02138, USA}

\author{Abraham Loeb}
\affil{Harvard-Smithsonian Center for Astrophysics, 60 Garden Street, Cambridge, MA 02138, USA}

\begin{abstract}

The possibility of Fast Radio Burst (FRB) emission being suppressed at low frequencies, resulting in a cutoff of the average rest-frame spectrum, has been raised as an explanation for the lack of detections at meter wavelengths. We examine propagation effects that could cause this suppression, and find that a low-frequency spectral cutoff may be generic regardless of the specific FRB emission mechanism. 
We then illustrate the effects of a low-frequency spectral cutoff on the statistics of FRBs, given a cosmological source population. The observed FRB rate peaks at a specific frequency under a variety of assumptions. Observations at lower frequencies are more sensitive to high-redshift events than observations above the maximal-rate frequency, and therefore result in more sharply broken fluence distributions. Our results suggest that the absence of low-frequency FRBs, and the differences between the Parkes and the Australian Square Kilometre Array (ASKAP) FRB samples, can be fully explained by suppressed low-frequency FRB emission. 

\end{abstract}


\section{Introduction}

In the past year, the range of frequencies over which fast radio bursts (FRBs) have been detected has been extended up to 8\,GHz \citep{gsp+18}, and down to 400\,MHz \citep{bcf+18}. FRB detection rates at frequencies between 400\,MHz and 1.8\,GHz are poised to improve by orders of magnitude with the advent of searches with the Canadian HI Intensity Mapping Experiment \citep[CHIME;][]{cab+18}, the Australian SKA Pathfinder \citep[ASKAP;][]{smb+18}, and the Deep Synoptic Array (Ravi et al., in preparation). It is therefore timely to assess the physical processes that shape FRB spectra, and their implications for FRB observations at different frequencies. 

Individual FRBs have entirely disparate spectra even within the typical observation bandwidths of a few hundred MHz around 1.4\,GHz \citep{lab+17,r18,msb+18}. Spectral structures on scales of $\sim100$\,kHz to tens of MHz are sometimes present even in the same FRB \citep{rsb+16}. However, the spectra of individual events may be shaped by stochastic processes, such as the intrinsic emission mechanism as in single pulses from pulsars \citep{kkg+03}, and time-variable diffractive scintillation effects \citep{cwh+17}. Our focus here is instead on the astrophysics of the {\em characteristic spectrum of the FRB phenomenon}, averaged over a large ensemble of events. As we shall show \citep[see also][]{vrh+16,fl17,me18b}, the characteristic FRB spectrum is a critical determinant of the observed-population demographics. 

As they involve coherent radio emission, FRBs are expected to be characterized by decreasing power-law spectra (indices $\alpha<0$) in the upper sections of their observed bandwidths.\footnote{We define the spectral index, $\alpha$, using a power-law flux density or fluence spectrum $\propto\nu^{\alpha}$.} This is the case for both normal and giant pulses\footnote{\citet{mtb+17} recently presented evidence for low-frequency flattening of the spectra of Crab giant pulses.} from pulsars \citep{kkg+03,mat+16}, which are most nearly analogous to FRBs among observed astronomical phenomena. This is also expected from models for the FRB emission mechanism \citep[][and references therein]{klb17}, which generally predict emission from coherent patches of particles with power-law energy distributions. However, pulsar observations \citep{klm+11,bkk+16,mkb+17,jvk+18} and predictions for the environments of FRB progenitors \citep{kon15,yzd16} combine to support the possibility of a spectral peak for FRBs, below which the characteristic FRB spectrum also decreases. 

The lack of FRB detections prior to the recent CHIME events at frequencies $\nu_{\rm obs}<700$\,MHz has led to the common belief that the FRB rate at these low frequencies is exceedingly low. A stacking analysis of 23 bandpass-calibrated FRB detections from ASKAP suggests a mean {\em observed} spectral index of $\alpha=-1.6^{+0.3}_{-0.2}$ between 1129--1465\,MHz \citep{msb+18}. Most efforts to observationally infer the characteristic FRB spectrum at lower frequencies have focused on comparing detection rates at different frequencies with the 1.4\,GHz rate set by the Parkes telescope \citep{kca+15,rbm+16,btb+16,cfb+17,ckj+17}. The non-detection in the 300-400\,MHz GBT North Celestial Cap survey \citep{ckj+17} is the most constraining, because the large survey time (84 days) was complemented by a better sensitivity (3.15\,Jy\,ms for a 5-ms FRB) than the 1.4\,GHz Parkes surveys. A characteristic spectral index of $\alpha>-0.3$ between the GBT and Parkes observing bands was derived, assuming a power-law FRB spectrum, even after accounting for pulse broadening caused by scattering in inhomogeneous plasma.  A complementary approach was adopted by \citet{sbm+18}, who presented non-detections of seven bright ASKAP FRBs between 170--200\,MHz with the Murchison Widefield Array (MWA). This result suggests a spectral index between the MWA and ASKAP bands that is shallower than $\alpha\sim-1$.  

Here we evaluate the astrophysical and observational implications of a peak or turnover frequency, $\nu_{\rm peak}$, in the characteristic rest-frame FRB spectrum. In \S\ref{sec:2}, we first show that several physical mechanisms can lead the existence of a low-frequency cutoff for FRBs. An observational constraint on $\nu_{\rm peak}$ can in turn be used to derive physical parameters of FRB progenitors and their environments. Second, besides shaping the frequency-dependent FRB rate, the presence of a low-frequency spectral cutoff modifies the fluence and redshift distributions of  FRBs observed at different frequencies. We demonstrate these effects in \S\ref{sec:3}. We discuss the observational consistencies and predictions of a low-frequency cutoff for FRBs in \S\ref{sec:4}, and conclude in \S\ref{sec:5}. In particular, we assert that the form of the characteristic FRB spectrum can independently explain other important observed features of the FRB population, such as the differences between the Parkes and ASKAP FRB samples \citep{smb+18,jem+18}. Throughout our discussion, we adopt the latest Planck cosmological parameters \citep{pc16}, with $H_{0}=67.7$\,km\,s$^{-1}$\,Mpc$^{-1}$, $\Omega_{b}=0.0486$, $\Omega_{M}=0.3089$, and $\Omega_{\Lambda}=0.6911$.

\section{Low-frequency modifications to FRBs} \label{sec:2}

\subsection{Physical mechanisms}

We consider propagation effects that can suppress the observed emission from FRBs at frequencies $\nu<\nu_{\rm peak}$. Our analysis makes no assumptions about the intrinsic FRB emission mechanism. We make a distinction between the effects we consider and those that decrease the observed signal to noise ratio (S/N) while preserving fluence, such as temporal broadening due to stochastic multi-path propagation through an inhomogeneous plasma. This is important because the fluence completeness thresholds of surveys can be well defined \citep[e.g.,][]{kp15} and controlled for in comparing observations at different frequencies. We also do not consider the mechanisms that shape the spectra of individual FRBs, because of the likely possibility that single bursts are realizations of a stochastic process with underlying stable ensemble characteristics.

\subsubsection{Plasma absorption}

Electromagnetic radiation cannot propagate through a plasma at frequencies 
\begin{equation}
    \nu\lesssim\nu_{p}=\left(\frac{n_{e}e^{2}}{\pi m_{e}}\right)^{1/2}\approx 90\left(\frac{n_{e}}{10^{8}\,{\rm cm}^{-3}}\right)^{1/2}\,{\rm MHz},
    \label{eqn:plasma}
\end{equation}
where $n_{e}$ is the electron density, $e$ is the electron charge, and $m_{e}$ is the electron mass. The plasma period, $\nu_{p}^{-1}$, corresponds to the characteristic timescale of Langmuir oscillations, or relaxations of density fluctuations in a plasma. If the electron temperature, $T_{e}$, is significant, such that the plasma is relativistic (i.e., $\sqrt{k_{B}T_{e}/m_{e}}\sim c$, where $k_{B}$ is Boltzmann's constant, corresponding to $T_{e}\gtrsim10^{9}$\,K), the plasma frequency is modified \citep{aap+75}.  

\subsubsection{Razin-Tsytovich effect}

The Razin-Tsytovich \citep[e.g.,][]{gs65} effect was first considered in terms of synchrotron emission from relativistic electrons within a thermal plasma. The effect of the refractive index in an electron plasma being less than unity is to widen the cone of relativistic beaming of the emission from individual electrons. This occurs for frequencies 
\begin{equation}
    \nu\lesssim\nu_{R}=\gamma\nu_{p}\approx 2.9 \left(\frac{n_{e}}{10^{8}\,{\rm cm}^{-3}}\right)\left(\frac{B}{1\,{\rm G}}\right)^{-1}\,{\rm GHz}
    \label{eqn:razin}
\end{equation}
where $B$ is the magnetic field strength in the emission region, and $\gamma$ is the electron Lorentz factor, which we eliminate by setting the emission frequency to $\gamma^{2}$ times the cyclotron gyrofrequency. We can express $B$ in terms of $n_{e}$ and $T_{e}$ for a thermal plasma assuming equipartition to find $\nu_{R}\approx 400(n_{e}/{\rm cm}^{-3})^{1/2}(T_{e}/{\rm K})^{-1/2}$\,MHz. This likely provides a lower limit on the characteristic frequency. This suppression was found to be applicable  to relativistic bremsstrahlung emission by \citet{m72}, and to coherent emission in pulsar magnetospheres by \citet{ab86}. As above, Equation~(\ref{eqn:razin}) applies only to a non-relativistic plasma. The Razin-Tsytovich effect has possibly been observed in solar radio bursts \citep{bc67}.

\subsubsection{Stimulated Raman scattering}

In the case of sources with high brightness temperatures, and therefore high emanent radiation energy densities, radio emission can also be Raman-scattered by Langmuir waves in dense plasma \citep{gk92,lb95}. Although the scattering minimally affects the spectra of isotropically radiating sources, sources that are beamed towards the observer are affected by the scattering of radiation away from the line of sight. The growth of the Langmuir oscillations in response to incident radiation is non-linear in the radiation energy density, and therefore in the brightness temperature $T_{b}$. Adopting a fiducial FRB brightness temperature of $10^{36}$\,K \citep{k14},\footnote{An FRB at a redshift of $z=0.5$ with a mean flux density of 1\,Jy, and a duration equal to the source light-crossing time of 1\,ms ($1/1.5$\,ms in the rest frame), corresponds to $T_{b}\approx10^{36}$\,K.} strong stimulated Raman scattering \citep{lb95} is in effect for 
\begin{eqnarray}
\label{eqn:srs}
&& \nu \lesssim 130\left(\frac{n_{e}}{{\rm cm}^{-3}}\right)^{1/2}\left(\frac{T_{e}}{{\rm K}}\right)^{-1/2}\,{\rm MHz}, \\
&& \frac{\nu}{4.5\times10^{-13}\,{\rm MHz}} \gtrsim \left(\frac{n_{e}}{{\rm cm}^{-3}}\right)^{1/2}\left(\frac{T_{b}}{10^{36}\,{\rm K}}\right)^{-1}\left(\frac{T_{e}}{{\rm K}}\right)^{-3/2}.
\end{eqnarray}
The ``weak'' case of stimulated Raman scattering identified by \citet{lb95} only modifies the latter of these two conditions. Hence, Equation~(\ref{eqn:srs}) is a strong constraint on the electron density and temperature of the medium surrounding FRB sources. 

\subsubsection{Induced Compton scattering}

In contrast to the case of stimulated Raman scattering, induced Compton scattering results in radio photons losing significant energy to thermal electrons in the presence of sufficient photon and electron densities \citep[e.g.,][]{cbr93}. These requirements are satisfied for a Thomson optical depth in excess of $0.02[T_{b}/(10^{12} \,{\rm K})]^{-1}$. For an assumed FRB brightness temperature of $T_{b}\approx10^{36}$\,K and a source radius $r_{\rm src} = 3\times10^{7}$\,cm, the requirement of a Thomson optical depth below the aforementioned value places an upper bound on the electron density of 
\begin{equation}
\label{eqn:ics}
n_{e}\lesssim10^{-9}\left(\frac{r_{\rm src}}{3\times10^{7}\,{\rm cm}}\right)\,{\rm cm}^{-3}.
\end{equation}
This is an extremely strong constraint, and possibly conservative given that FRB emission is likely beamed. We note that in the regime where both stimulated-Raman and induced-Compton scattering are active, Raman scattering likely dominates \citep{lb95}. 

One might infer that the high brightness temperature of pulsar emission implies that induced Compton scattering is important in that scenario as well. This has been shown not to be the case \citep[e.g.,][]{wr78}, because of the relativistic nature of pulsar winds. Thermal plasma in supernova remnants surrounding young pulsars is also not significant. Consider the highest brightness temperature reported for a Crab giant pulse of $10^{41}$\,K for an estimated emission region size of 10\,cm \citep{he07}. For $r_{\rm src}\sim10^{16}$\,cm in the inner regions of supernova remnants, we require $n_{e}\gtrsim3\times10^{7}$\,cm$^{-3}$, in contrast to the requirement of $n_{e}\gtrsim0.3$\,cm$^{-3}$ for FRBs. 

The spectral distortions caused by induced Compton scattering mildly suppress the spectral energy distributions of sources at all frequencies where the brightness temperature is sufficiently high. Although the specific distortions are dependent on the geometry of the source \citep{cbr93}, induced Compton scattering for a spherical source permeated by or embedded within a spherical cloud of thermal plasma will have a spectral index of $\alpha=1$ for a flat or positively sloped input spectrum, and $\alpha=1-\alpha'/2$ for an input spectrum with an input spectral index of $-\alpha'$ \citep{s71}.  For a non-thermal source, the brightness temperature and therefore the susceptibility to induced Compton scattering varies with the emission frequency. As the actual brightness temperatures of FRBs are highly uncertain, we simply adopt the estimate of $T_{b}\approx10^{36}$\,K at emission frequencies around 1\,GHz.

\subsubsection{Free-free absorption}

A thermal plasma surrounding an FRB source will also result in free-free absorption of the incident radiation. This effect has been previously considered as a necessary constraint to overcome in progenitor models involving young neutron stars \citep[e.g.,][]{kon15}. Defining a characteristic frequency as that below which the free-free optical depth is greater than unity, free-free absorption results in attenuation for frequencies
\begin{equation}
\label{eqn:ff}
    \nu\lesssim 300(T_{e}/{\rm K})^{-0.64}{\rm EM}^{0.48}\,{\rm MHz},
\end{equation}
where the emission measure (EM) of the absorbing plasma is in standard units of pc\,cm$^{-6}$ \citep{d11}.

\subsubsection{Propagation through inhomogeneous plasma}

We have noted that temporal broadening of FRBs due to multi-path propagation through inhomogeneous plasma is expected to conserve fluence, and we therefore do not consider its effect on surveys with known fluence-completeness thresholds. However, it is additionally possible that FRBs are strongly magnified by AU-scale plasma lenses in their host galaxies \citep{cwh+17}, or by the effects of constructive interference of rays propagating along multiple paths \citep[e.g.,][]{rsb+16}. Although lensing caustics impose a rich frequency structure on the magnified input spectrum, with spectral peaks with $\sim0.1$--1\,GHz widths magnified by up to factors of $\sim100$, the effects can be quite broadband despite the chromaticity of the plasma refractive index. Interference maxima resulting from strong scattering in FRB host galaxies, with characteristic ray delays $\tau_{d}$ producing spectral peaks with widths $\sim1/(2\pi\tau_{d})$ and an exponential intensity distribution, are also difficult to relate to low-frequency FRB suppression, because of the strong reduction of $\tau_{d}$ with redshift \citep[stronger than $(1+z)^{3}$;][]{mk13}. However, as noted by \citet{msb+18}, a combination of angular broadening and interference due to scattering in both FRB host galaxies and intervening systems may conspire to magnify FRBs only at GHz frequencies. 

\subsubsection{Lessons from Galactic pulsars}

Pulsar emission is the closest known analog to FRBs. Pulsars often have peaks or breaks at frequencies $\nu_{\rm peak}$ between 0.1--few GHz \citep{klm+11,bkk+16,jvk+18}. Multiple mechanisms likely define these critical frequencies. For example, although a positive correlation may exist between $\nu_{\rm peak}$ and DM, and some GHz-peaked spectrum pulsars are viewed along particularly dense sightlines, free-free absorption effects are difficult to disentangle from increased scattering at low frequencies. In addition, young pulsars tend to have flatter radio spectra, and higher values of $\nu_{\rm peak}$, than older pulsars. Although this may suggest that pulsar spectra depend sensitively on the magnetospheric properties, this trend may also be a consequence of young pulsars residing closer to their extreme birth environments. Third, as persuasively argued by \citet{jvk+18}, our knowledge of pulsar spectra generally improves upon closer inspection; the best studied pulsars, observed over the largest number of epochs to mitigate the effects of scattering and possible intrinsic variations, have complex broadband spectra that cannot be characterized by single power laws. 

\subsection{Application to FRB progenitor environments}

\begin{figure}
    \centering
    \includegraphics[width=0.47\textwidth]{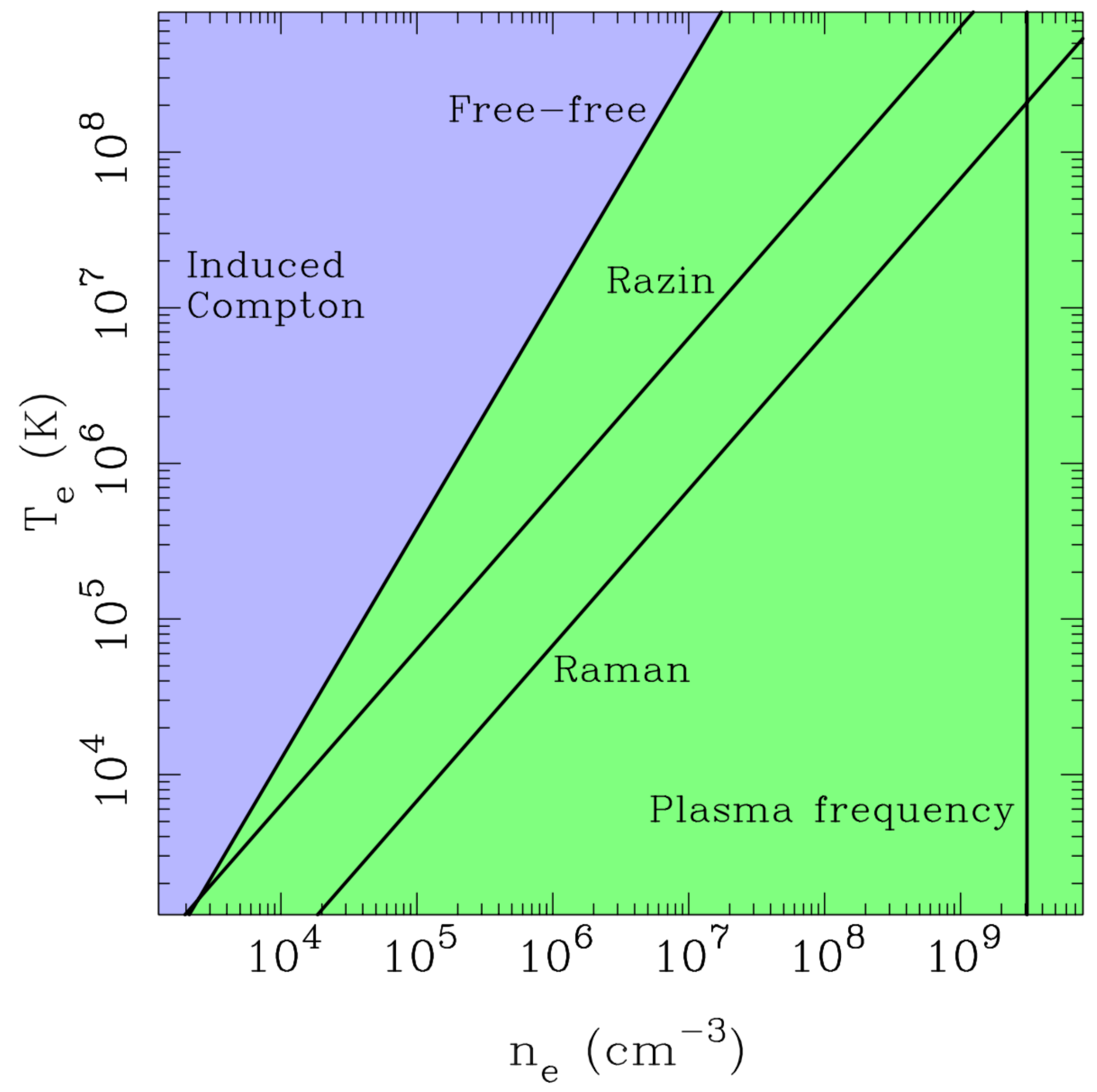}
    \caption{Loci in $T_{e}-n_{e}$ space derived from different FRB absorption/suppression mechanisms assuming $\nu_{\rm peak}=500$\,MHz. Radiation is suppressed for $\nu<500$\,MHz, which is generally the case for higher values of $n_{e}$ and lower values of $T_{e}$. As labeled, the lines indicate constraints from absorption below the plasma frequency (Equation~\ref{eqn:plasma}), Razin-Tsytovich suppression (Equation~\ref{eqn:razin}) assuming a plasma beta of unity, stimulated Raman scattering (Equation~\ref{eqn:srs}), and free-free absorption in a constant-density nebula of 0.1\,pc radius (Equation~\ref{eqn:ff}). The green shaded region illustrates the possibly excluded region assuming that the free-free optical depth at 500\,MHz is less than unity. The purple shaded region labeled as induced Compton scattering corresponds to radiation below $\nu_{\rm peak}\sim1$\,GHz being suppressed for ionized-nebula sizes $r_{\rm src}\lesssim10$\,pc (Equation~\ref{eqn:ics}). For example, setting $r_{\rm src}=0.1$\,pc only requires a density of $n_{e}>10$\,cm$^{-3}$ for induced Compton scattering to occur.}
    \label{fig:1}
\end{figure}

Some of the physical mechanisms identified above can plausibly result in a characteristic low-frequency spectral cutoff for FRBs. Additionally, if the frequency of such a cutoff can be determined, the properties of the immediate progenitor environments of FRBs can be constrained. Figure~\ref{fig:1} illustrates necessary values of the density and temperature of a thermal, non-relativistic plasma surrounding (and permeating, in the case of Razin-Tsytovich suppression) FRB sources, for a characteristic frequency of $\nu_{\rm peak}=500$\,MHz. The Razin-Tsytovich effect (Equation~\ref{eqn:razin}) is evaluated assuming equipartition between the thermal and magnetic energy densities in the plasma, and the EM for the free-free absorption constraint (Equation~\ref{eqn:ff}) is  evaluated for a constant-density nebula with a 0.1\,pc radius. 

The required combinations of $T_{e}$ and $n_{e}$ for $\nu_{\rm peak}\leq500$\,MHz are plausible for Razin-Tsytovich suppression and stimulated Raman scattering, and for a $\sim0.1$\,pc ionized nebula in the case of free-free absorption. In practice, however, free-free absorption may not be the preferred low-frequency suppression mechanism because of the requirement for large DM contributions from the host environment. Absorption below the plasma frequency is less likely. Induced Compton scattering will occur for any combination of $T_{e}$ and $n_{e}$ in Figure~\ref{fig:1} for ionized nebula sizes $r_{\rm src}\lesssim10$\,pc. However, in comparison with the other suppression/absorption mechanisms, induced Compton scattering less strongly distorts the observed low-frequency spectrum. 

The results in Figure~\ref{fig:1} can be compared with expectations for different astrophysical environments that could host FRB progenitors. FRB progenitor models generally posit a compact object as the engine, most often invoking a highly magnetized neutron star \citep[for a summary of FRB progenitor models, see][]{pww+18}. Young and massive compact objects are likely surrounded by dense, hot plasma. The environments of the repeating FRB \citep{msh+18} and some non-repeating events \citep[e.g.,][]{mls+15} are likely dense and magnetized, corresponding to young supernova remnants or the surrounds of supermassive black holes (SMBHs).  
\begin{description}
\item[Young neutron star in its supernova remnant] Extreme emission from young pulsars \citep[e.g.,][]{cw16} or magnetars \citep[e.g.,][]{mbm17} forms a leading model for FRBs. Although older supernova remnants like the Crab nebula self-evidently do not result in absorption of radiation at $\nu\gtrsim100$\,MHz, the environments of neutron stars younger than $\sim100$\,yr are more extreme.  \citet{mmb+18} computed photoionization models of superluminous supernova remnants that potentially host nascent FRB-emitting magnetars, finding $T_{e}\sim10^{6}-10^{7}$\,K in the central (post-shock) regions where number densities $n_{e}\gtrsim10^{5}$\,cm$^{-3}$ are expected \citep{mbm17}. Significant pre-explosion mass loss is also inferred for some engine-driven core collapse supernovae, resulting in post-shock densities of $n_{e}\gtrsim10^{6}$\,cm$^{-3}$ within $10^{14}$\,cm of the center \citep[for a compilation, see][]{hpr+18}.
\item[Pulsar wind] Pulsars lose most of their spin-down energy to winds of relativistic particles. Assuming a bulk Lorentz factor of the wind of $\sim10^{2}$ just beyond the light cylinder \citep[][and references therein]{gs06}, negligible magnetization of the wind, and a standard neutron-star radius of 10\,km, the particle density in the wind at a radius of 1000\,km is 
\begin{equation}
    n_{\rm pw} \sim 10^{16}\left(\frac{B_{p}}{10^{14}\,{\rm G}}\right)^{2}\left(\frac{P}{1\,{\rm s}}\right)^{-4},
\end{equation}
where $B_{p}$ is the polar magnetic field strength, and $P$ is the spin period. Evaluating the absorption properties of such a wind is difficult because of its relativistic nature \citep[e.g.,][]{wr78}. The wind is also likely magnetically dominated at this location, transitioning to a particle-dominated outflow only near the termination shock, where bulk Lorentz factors of $\sim10^{6}$ are in fact inferred. 
\item[Supermassive black holes] Wildly variable assessments exist in the literature of the plasma environments surrounding SMBHs. \citet{abg+16} collated inferences from radio outflows/jets from tidal disruption events to find typical densities of $10^{4}-10^{5}$\,cm$^{-3}$ at radii of $10^{15}$\,cm. As an example, \citet{b52} accretion onto a supermassive black hole from a $10^{6}$\,K interstellar medium (ISM) will result in a density at $10^{15}$\,cm that is $\sim10^{4}(M_{\rm BH}/10^{6}M_{\odot})$ times higher than the ISM density, where $M_{\rm BH}$ is the SMBH mass.  
\end{description}

In summary, a selection of plausible FRB progenitor environments can result in the suppression of emission below $\nu_{\rm peak}\sim1$\,GHz. We stress that all mechanisms, including Razin-Tsytovich suppression, stimulated Raman scattering, and induced Compton scattering, need to be accounted for in FRB progenitor models. We now turn our attention to the observational consequences of a characteristic low-frequency spectral cutoff for FRBs, addressing the consistency of this model with current observations, and predictions of this model that may allow $\nu_{\rm peak}$ to be measured.

\section{The observed fluence and redshift distributions} \label{sec:3}

We use a straightforward fiducial model for the characteristic FRB fluence spectrum and luminosity function to demonstrate the effects of a characteristic rest-frame low-frequency spectral cutoff for FRBs. Our analysis assumes a cosmological FRB population, such that FRBs originate from redshifts wherein the extragalactic DMs are dominated by propagation through the circum- and inter-galactic medium. A growing compilation of observations supports this scenario. For example, the repeating FRB\,121102 is observed at a large extragalactic distance \citep{tbc+17}, FRB\,150807 had no nearby host galaxies within its localization region \citep{rsb+16}, and evidence exists for a relation between fluence and DM consistent with a cosmological population \citep{smb+18}. 

We adopt a two-component power law for the fluence spectrum:
\begin{eqnarray}
F(\nu_r) &=& F_{0}(\nu_r/\nu_{\rm peak})^{\alpha},\,\nu_r\geq\nu_{\rm peak} \\
&=& F_{0}(\nu_r/\nu_{\rm peak})^{\beta},\,\nu_r<\nu_{\rm peak},
\end{eqnarray}
where $\nu_r$ is the rest frequency, and $\alpha>0$ and $\beta<0$ are the two spectral indices. As noted above, we consider this to be a ``central'' FRB spectrum, which the individual rest-frame spectra of FRBs tend towards on average. Evidence for the existence of such an FRB spectrum was recently provided by \citet{msb+18}, who showed that the calibrated spectra of 23 ASKAP FRBs tended towards a central value upon averaging, rather than having infinite variance. 

The values of $\alpha$ and $\beta$ are difficult to specify a priori. Specifically, the value of $\alpha$ is set by the intrinsic FRB emission mechanism, and $\beta$ is set by the low-frequency suppression mechanism. Induced Compton scattering will result in $\beta=1-\alpha/2$ for $\alpha\leq0$, free-free absorption will result in $\beta=2.1$, and the remaining mechanisms considered in \S\ref{sec:2} will result in much steeper cutoffs. For the purposes of our demonstration in this section, we assume $\alpha=-1.8$ \citep{msb+18}, and consider illustrative values of $\beta=2.1$ and $\beta=\infty$.  

Let $n_{\rm ref}(>F)$ be the observed number of FRBs per unit time at a frequency $\nu_{\rm obs}$, per comoving volume element at a fiducial redshift $z_{\rm ref}$, a fiducial observing frequency $\nu_{\rm ref}$, above an observed fluence $F$. We have little guidance in what functional form to adopt for $n_{\rm ref}(>F)$. For consistency with the observations of single pulses from pulsars \citep{mmm+18}, but not the extreme case of giant pulse emission, we adopt a log-normal form for the differential FRB counts, $\frac{d}{dF} n_{\rm ref}(>F)$, with mean $\ln F_{\rm ref}$ and standard deviation $\sigma_{\rm ref}^{2}$. Another possibility would have been the Weibull distribution, which describes the statistics of maximal extreme values. A power law distribution has the disadvantage of having more free parameters, including the arbitrary choices of low- and high-fluence cutoffs \citep[cf.][]{fl17,me18b}. Again for the purposes of demonstration, we adopt fiducial values of $\nu_{\rm ref}=1$\,GHz, and $\sigma_{\rm ref}=0.3$\,dex; we do not need to specify $F_{\rm ref}$. We adopt an arbitrary normalization for the total volumetric rate, and only present relative quantities.

The equivalent quantity to $n_{\rm ref}(>F)$ at a frequency $\nu$ and a redshift $z$ is given by 
\begin{equation}
n(z,>F) = R(z)\frac{(1+z_{\rm ref})}{(1+z)}n_{\rm ref}(>F'),
\label{eqn:unit}
\end{equation}
where $R(z)$ captures any cosmic evolution in rest-frame FRB volumetric rate, and the effective fluence is given by 
\begin{equation}
F' = F\left[\frac{D_{L}(z_{\rm ref})}{D_{L}(z)}\right]^{-2}\frac{F[(1+z_{\rm ref})\nu_{\rm ref}]}{F[(1+z)\nu_{\rm obs}]}\frac{(1+z_{\rm ref})}{(1+z)}.
\end{equation}
Here, the $\left[\frac{D_{L}(z_{\rm ref})}{D_{L}(z)}\right]^{-2}$ factor captures the evolution of flux density with luminosity distance $D_{L}$, the $\frac{F[(1+z_{\rm ref})\nu_{\rm ref}]}{F[(1+z)\nu]}$ factor corrects the observed fluence for the redshifted frequency \citep[which we refer to as the $K$-correction;][]{hbb+02}, and the $\frac{(1+z_{\rm ref})}{(1+z)}$ factor corrects the observed fluence for the dilated duration. The number of observed FRBs above a fluence $F$, at a frequency $\nu_{\rm obs}$, is given by the redshift integral over Equation~(\ref{eqn:unit}):
\begin{equation}
\label{eqn:lnls}
N(>F) = \int_{0}^{\infty}n(z,>F)\frac{4\pi d^{2}V_{C}}{d\Omega dz}dz,
\end{equation}
where $d^{2}V_{C}/(d\Omega dz)$ is the standard comoving volume element. The redshift distribution of FRBs observed above a fluence $F$, at a frequency $\nu$, is then 
\begin{equation}
\label{eqn:dndz}
\frac{d}{dz}N(>F) = n(z,>F)\frac{4\pi d^{2}V_{C}}{d\Omega dz}.
\end{equation}

\subsection{Frequency-dependent detection rate}

\begin{figure}[t!]
    \centering
    \includegraphics[width=0.48\textwidth,trim=0cm 0cm 0.5cm 1cm,clip]{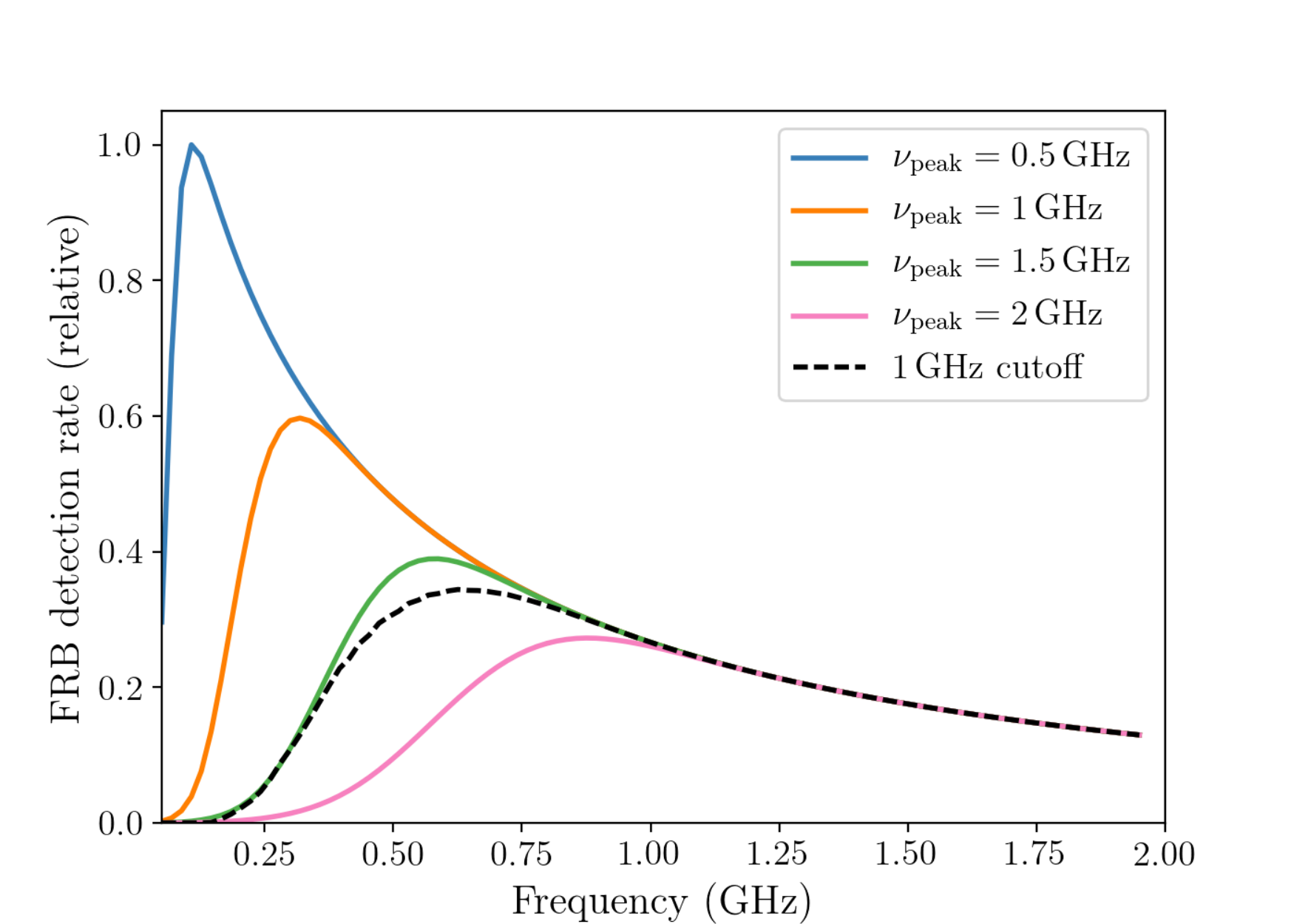}
    \caption{The relative FRB detection rates at different frequencies. We show results for an FRB spectrum described by $\alpha=-1.8$ and $\beta=2.1$ for four values of $\nu_{\rm peak}$ (solid curves), and for an FRB spectrum with $\alpha=-1.8$ and a sharp cutoff ($\beta=\infty$) below $\nu_{\rm peak}=1$\,GHz (dashed curve). We consider We assume FRB detection thresholds at each frequency that are equivalent to $F_{\rm ref}$, and set the maximum FRB redshift to 6. We set $\nu_{\rm ref}=1$\,GHz and $\sigma_{\rm ref}=0.3$\,dex to describe the FRB luminosity function.}
    \label{fig:2}
\end{figure}

We begin by illustrating the effects of a characteristic low-frequency spectral cutoff for FRBs by calculating the detection rates corresponding to various values of $\nu_{\rm peak}$ for surveys at different frequencies. We assume that all surveys have identical fluence thresholds, corresponding to $F_{\rm ref}$, and that all surveys search over the same DM ranges. This latter assumption aids in evaluating the redshift integral in Equation~(\ref{eqn:lnls}) by setting a maximum redshift. We assume a maximum DM of 6000\,pc\,cm$^{-3}$, which we assume originates predominantly in the circum- and intergalactic medium, and which therefore crudely corresponds to a redshift of 6 \citep[e.g.,][]{i03}. Finally, we assume no redshift evolution of the volumetric FRB rate (constant $R(z)$ in Equation~\ref{eqn:unit}).  The results are shown in Figure~\ref{fig:2}.

For an FRB spectrum described by $\alpha=-1.8$ and $\beta=2.1$, the correlation between $\nu_{\rm peak}$ and the frequency at which the detection rate peaks is evident. At high frequencies, all cases approach the same curve as all detectable sources are observed at rest-frequencies above $\nu_{\rm peak}$. The lower values of $\nu_{\rm peak}$ result in higher detection rates at low frequencies because sources are observed above $\nu_{\rm peak}$ at lower redshifts, where more of the population is accessible.  The FRB spectrum with a sharp cutoff below $\nu_{\rm peak}=1$\,GHz has a lower detection rate at low frequencies than the former case with the same $\nu_{\rm peak}$, because sources are no longer amplified into the detection volume by the negative $K$-correction. 

\subsection{Observed redshift distributions}

\begin{figure*}[t!]
    \centering
    \includegraphics[width=0.48\textwidth,trim=0cm 0cm 1cm 1cm,clip]{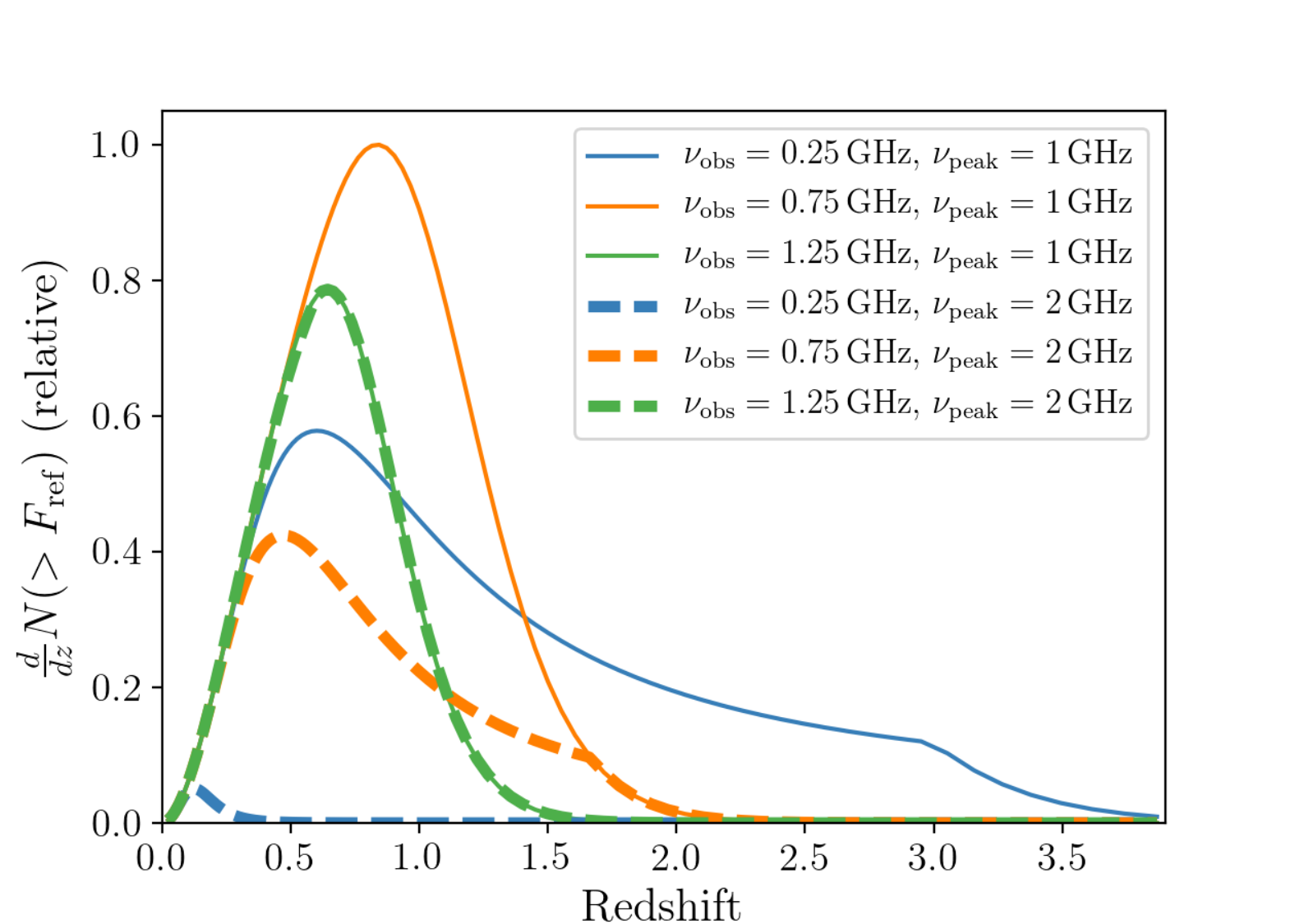}
    \includegraphics[width=0.48\textwidth,trim=0cm 0cm 1cm 1cm,clip]{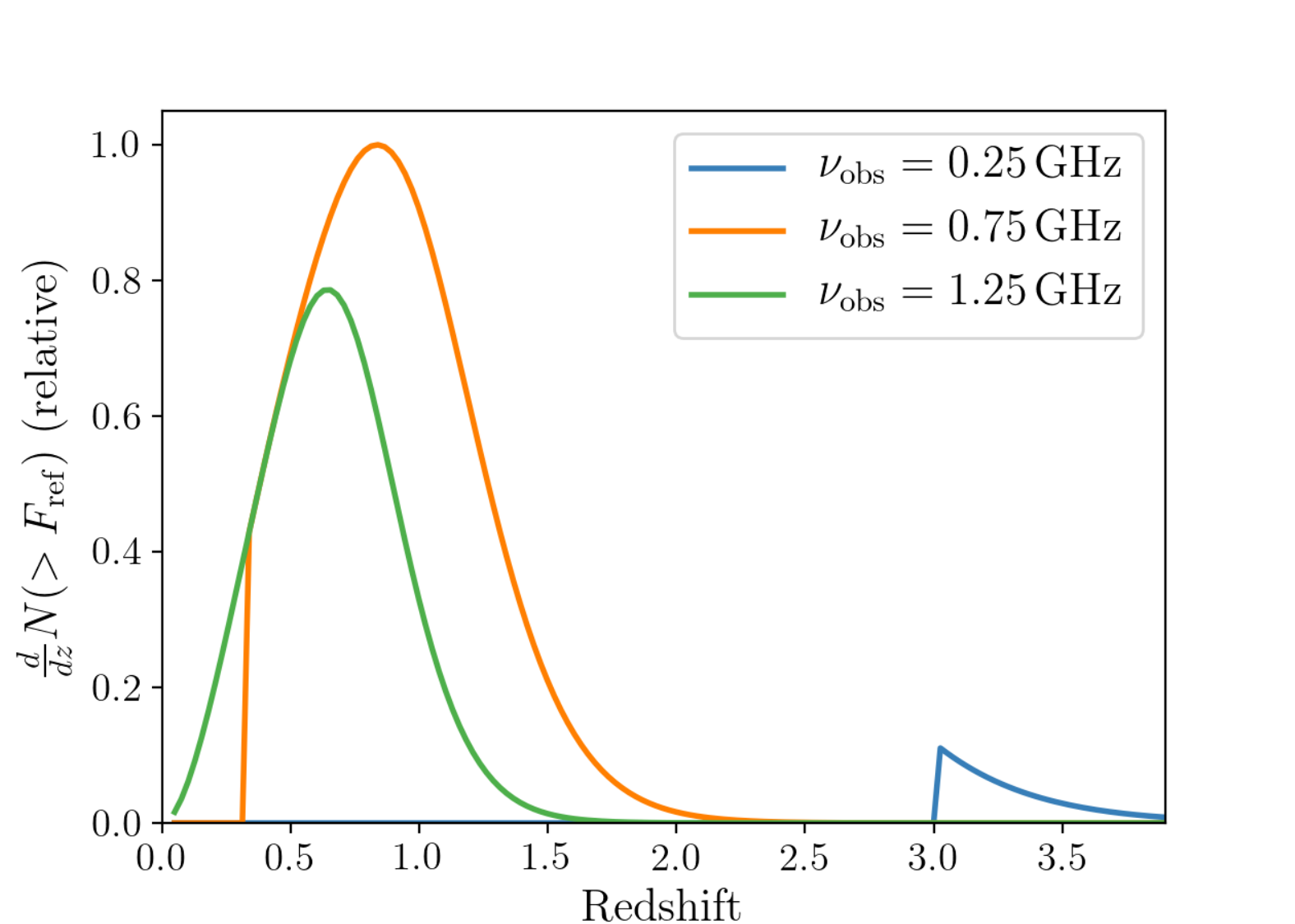}
    \caption{{\em Left panel:} the differential FRB rate at different redshifts (Equation~\ref{eqn:dndz}) for observations at $\nu_{\rm obs}=0.25,\,0.75,\,1.25$\,GHz, and $\nu_{\rm peak}=1,\,2$\,GHz (solid and dashed lines respectively). An FRB spectrum described by $\alpha=-1.8$ and $\beta=2.1$ is assumed, and the luminosity function is as above (e.g., Figure~\ref{fig:2}). {\em Right panel:} same as the left panel, but for an FRB spectrum with $\alpha=-1.8$ and a sharp cutoff ($\beta=\infty$) below $\nu_{\rm peak}=1$\,GHz.}
    \label{fig:3}
\end{figure*}

We next illustrate the redshift distributions of FRBs observed at different frequencies relative to the frequency with the peak rate. We again assume a common fluence threshold of $F_{\rm ref}$, and a constant $R(z)$ in Equation~(\ref{eqn:unit}). The differential redshift distribution of FRBs observed above a fluence $F$ at a frequency $\nu$ is given by Equation~(\ref{eqn:dndz}). We consider observations at frequencies below ($\nu_{\rm obs}=0.25$\,GHz), around ($\nu_{\rm obs}=0.75$\,GHz), and above ($\nu_{\rm obs}=1.25$\,GHz) the frequencies with the peak FRB rates for $\nu_{\rm peak}=1$\,GHz and $\nu_{\rm peak}=2$\,GHz. Results are shown in Figure~\ref{fig:3} for both FRB spectral models discussed above.

Observations at low frequencies tend to be more sensitive to high-redshift FRBs than observations at higher frequencies. This is because of the negative $K$-correction, which results in more distant FRBs being brighter than expected from the $D_{L}^{2}$ law as they are observed closer to their spectral peaks. The breaks evident in the left panel of Figure~\ref{fig:3} for the $\nu_{\rm obs}=0.25$\,GHz and $\nu_{\rm peak}=1$\,GHz curve, and the $\nu_{\rm obs}=0.75$\,GHz and $\nu_{\rm peak}=2$\,GHz curve, correspond to the redshifts where $\nu_{\rm peak}=(1+z)\nu_{\rm obs}$ (higher redshift FRBs are all observed above their spectral peaks). As shown in the right panel of Figure~\ref{fig:3}, a sharper characteristic low-frequency spectral cutoff will result in a more pronounced bias towards high-redshift FRBs for low-frequency observations, together with a potentially associated drop in overall detection rate for appropriate luminosity functions and detection thresholds. This is because FRBs only become detectable at low frequencies when their emission redshifts into view. 

\subsection{Observed fluence distributions}

\begin{figure}[t]
    \centering
    \includegraphics[width=0.48\textwidth,trim=0cm 0cm 1cm 1cm,clip]{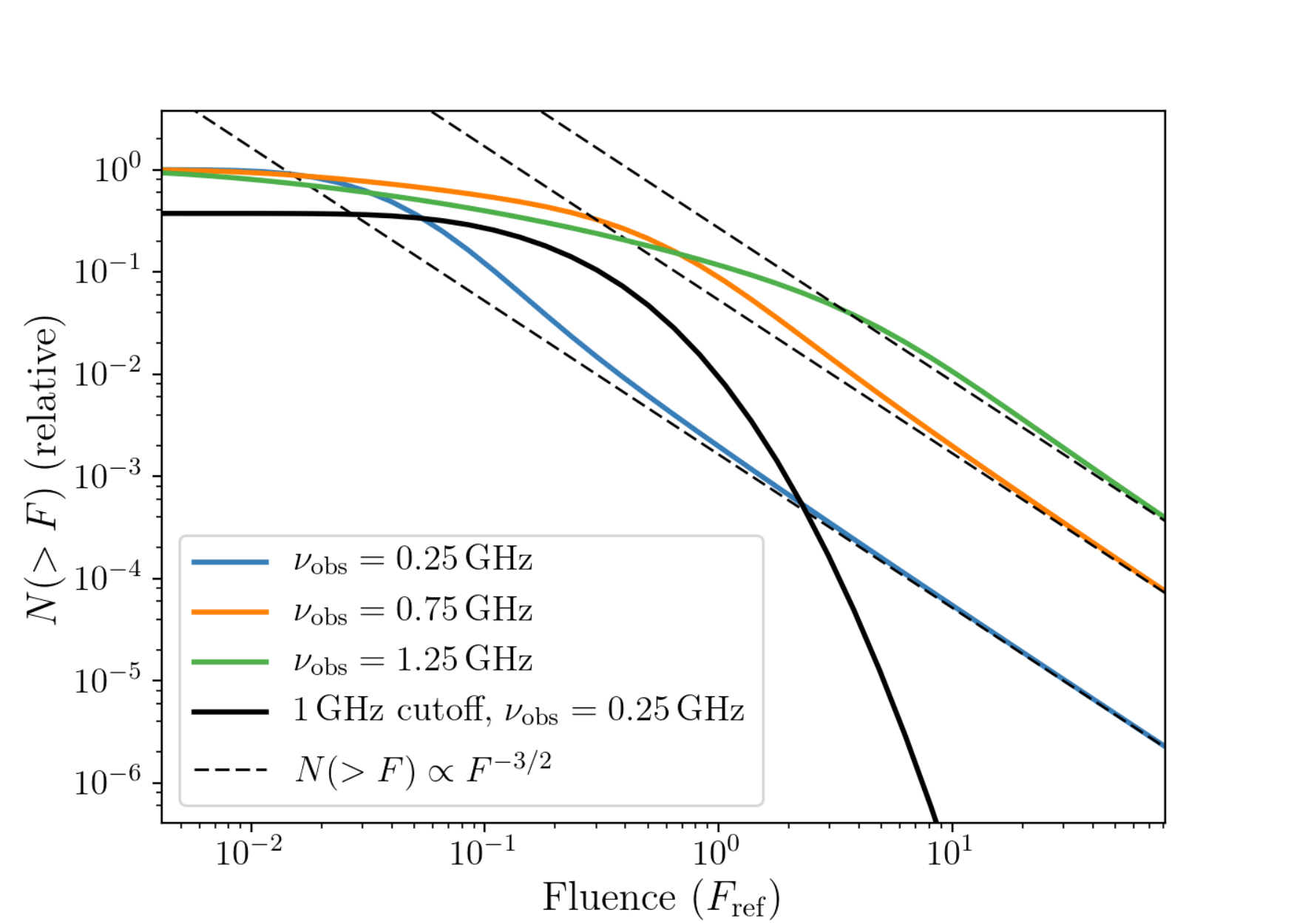}
    \caption{The observed integral FRB fluence distributions (Equation~\ref{eqn:lnls}) for $\nu_{\rm peak}=2$\,GHz at $\nu_{\rm obs}=0.25,\,0.75,\,1.25$\,GHz and a [$\alpha=-1.8$, $\beta=2.1$] FRB spectrum (blue, orange, and green curves), and for $\nu_{\rm peak}=1$\,GHz and $\nu_{\rm obs}=0.25$\,GHz for the sharply cutoff spectrum considered above. The assumptions regarding the FRB luminosity function are the same as above (e.g., Figures~\ref{fig:2}, \ref{fig:3}). The dashed black lines show indicative fluence distributions corresponding to uniformly distributed sources in Euclidean space ($N(>F)\propto F^{-3/2}$).}
    \label{fig:4}
\end{figure}

Finally, we illustrate the fluence distributions of FRB samples observed at different frequencies relative to the frequency with the peak rate (Equation~\ref{eqn:lnls}). For an FRB spectrum described by $\alpha=-1.8$ and $\beta=2.1$, we calculate the fluence distributions for $\nu_{\rm peak}=2$\,GHz at $\nu_{\rm obs}=0.25,\,0.75,\,1.25$\,GHz. We also calculate the fluence distribution for the cutoff spectrum below $\nu_{\rm peak}=1$\,GHz, at $\nu_{\rm obs}=0.25$\,GHz. Thus, the fluence distributions we evaluate correspond exactly to the redshift distributions shown in Figure~\ref{fig:3}.  The results are shown in Figure~\ref{fig:4}.

The expectation for a uniformly distributed source population in Euclidean space, which closely corresponds to a nearby cosmological source population, is a relation $N(>F)\propto F^{-3/2}$. At the high-fluence end of the fluence distributions for the [$\alpha=-1.8$, $\beta=2.1$] spectral model in Figure~\ref{fig:4}, the curves (will) asymptote to this relation, because of the bounded nature of the assumed FRB luminosity function. The fluence distributions for values of $\nu_{\rm obs}$ below the frequency with the peak FRB rate (which would be $\nu_{\rm obs}\approx1$\,GHz; Figure~\ref{fig:2}) have a relative excess of faint events, or equivalently a paucity of bright events. Indeed, these fluence distributions are steeper than the fiducial $F^{-3/2}$ law in portions of their domains, approaching $F^{-2}$. This is because of the excess of higher-redshift, fainter events observed at these low frequencies, indicated in Figure~\ref{fig:3}. The flatness of the fluence distribution at the higher frequency of $\nu_{\rm obs}=1.25$\,GHz is due to a combination of the positive $K$-correction caused by the negatively sloped FRB spectrum observed at this frequency, and cosmic evolution of the comoving volume element. 

The fluence distribution at $\nu_{\rm obs}=0.25$\,GHz for the spectral model with a sharp cutoff below $\nu_{\rm peak}=1$\,GHz is quite different to the above cases. High-fluence, nearby events are no longer present because of the lack of low-redshift FRBs radiating at the observing frequency; the $F^{-3/2}$ behavior at high fluences is no longer evident. Low-fluence events are also suppressed by the more stringent detectability constraint on the most distant FRBs. In the case of a sharp spectral cutoff, sensitive low-frequency observations will predominantly detect high-redshift events \citep[see also][]{fl17}. 

\section{Discussion} \label{sec:4}

\subsection{Postdictions}

We have demonstrated the possible effects of a characteristic low-frequency spectral cutoff for FRBs on some of the observed population statistics. The precise nature of the frequency-dependent rate, and the redshift and fluence distributions, depend sensitively on the FRB luminosity function, and the form of the characteristic rest-frame spectrum. This is the case even in the absence of cosmic evolution in the population. Further, the frequency of the spectral cutoff, although plausibly at GHz frequencies, itself depends sensitively on the nature of FRB environments. These quantities are all poorly constrained at present. We therefore do not attempt to directly model the observed FRB population in this paper. 

However, it is clear that a low-frequency spectral cutoff can explain the tight constraints on the characteristic FRB spectral index between the Parkes/ASKAP frequency bands, and the GBT 350\,MHz \citep{ckj+17} and MWA 170--200\,MHz surveys \citep{sbm+18}. For example, the results shown in Figure~\ref{fig:2} demonstrate that equivalent detection rates for surveys with identical fluence thresholds can be obtained at $350$\,MHz and $1.4$\,GHz for a rest-frame turnover frequency $\nu_{\rm peak}\sim1$\,GHz. The general requirement for a frequency dependent FRB rate, with a maximal-rate frequency that is correlated with $\nu_{\rm peak}$, is that the FRB luminosity function has a form such that brighter FRBs are generally observed at lower luminosity distances. This can be achieved in several ways besides with the log-normal form for the luminosity function we assume above, such as with a very steeply negative or cutoff power law. This may indeed be the case in reality if FRB DMs are a proxy for cosmological distance \citep{smb+18}. 

A characteristic low-frequency spectral cutoff can also account for the differences between the Parkes and ASKAP FRB samples. The Parkes sample appears to have a burst fluence distribution that is flatter than the Euclidean-space expectation \citep[approximately $\propto F^{-1}$;][]{vrh+16,jem+18}, whereas the ASKAP sample appears to have a much steeper fluence distribution \citep[approximately $\propto F^{-2.2}$;][]{jem+18}.\footnote{There has been substantial uncertainty regarding the Parkes FRB fluence distribution \citep[e.g.,][]{ocp16,vrh+16,me18}. The continued detections of Parkes FRBs in multiple beams of the 13-beam array \citep[e.g.,][]{osj+18}, together with the re-assessment of the \citet{me18} technique by \citet{jem+18}, points towards a growing consensus in favor of a flat fluence distribution amongst Parkes FRBs.} The ASKAP detection rate is also lower than expected from an extrapolation of the Parkes fluence distribution. A model for the FRB fluence distribution wherein a break exists between the Parkes and ASKAP fluence-detection regimes is thus favored \citep[see also][]{abb+17}. This was attributed by \citet{smb+18} and \citet{jem+18} to a cosmologically evolving population, with a higher volumetric rate at higher redshifts. However, a similar FRB fluence distribution, containing portions that are both flatter and steeper than the fiducial $F^{-3/2}$ law, can also be generated by observations at low frequencies relative to a rest-frame cutoff (e.g., Figure~\ref{fig:4}, $\nu_{\rm obs}=0.25$\,GHz curves in both panels). In this model, the higher-fluence ASKAP events are observed typically from lower redshifts than the Parkes events \citep[consistent with their lower DMs;][]{smb+18}, and the Parkes rate at higher redshifts is boosted by a negative $K$-correction. We stress that the analysis in \S\ref{sec:3} is not intended to present a quantitative prediction of the cutoff frequency $\nu_{\rm peak}$ in this scenario, although it is likely to be $\gtrsim2$\,GHz. We also remark that the differing spectral properties of the Parkes and ASKAP FRBs \citep{msb+18} may also result from different regions of the characteristic rest-frame FRB spectrum being observed by the two surveys. 

The steeply negative spectral index of $\alpha=-1.6^{+0.3}_{-0.2}$ between 1129--1465\,MHz measured by \citet{msb+18} for the ASKAP FRB sample  presents a challenge to our proposed scenario. Given this observation, it is difficult to simultaneously explain the GBT/MWA non-detections and the differing properties of the Parkes and ASKAP samples using a single characteristic low-frequency cutoff. It is possible that ASKAP does indeed observe FRBs at rest-frequencies $\nu>\nu_{\rm peak}$, in which case cosmic evolution may be required to explain the differing source counts of the Parkes and ASKAP samples. Alternatively, it is possible that this measurement is erroneous, in which case the tension with a concordance scenario of a single $\nu_{\rm peak}$ is removed. For example, accurate estimates of the FRB rest-frame spectrum may rely on redshift corrections being applied, which would be the case if some FRBs are observed near $\nu_{\rm peak}$ and hence do not exhibit power-law spectra \citep[as assumed by][]{msb+18}.  

A further challenge to our hypothesis of a characteristic {\em rest-frame} spectral cutoff is the possibility that FRBs are viewed along a wide variety of sightlines. For example, FRB\,110523 was potentially Faraday-rotated by magnetic fields slightly in excess of the typical Milky Way ISM, but was also more strongly scattered by the circum-burst medium \citep{mls+15}. FRB\,150807, on the other hand, appeared to be neither scattered nor Faraday-rotated by a potential host-galaxy ISM comparable to that of the Milky Way \citep{rsb+16}. Although the repeating FRB\,121102 may not share a progenitor with the remainder of the population \citep[e.g.,][]{r18,smb+18}, its environment appears significantly different again, hosting $\gtrsim$\,mG magnetic fields but potentially not strongly scattering the bursts \citep[e.g.,][]{msh+18}. A variety of host environments may result in a range of values of $\nu_{\rm peak}$. However, it is not clear that current data strongly constrain the plasma environments immediately surrounding FRB sources on sub-parsec scales, because of the low column-densities (of order unity) involved, and the suppression of scattering owing to the extreme geometry. Both the stimulated-Raman and induced-Compton scattering mechanisms only require significant plasma densities rather than column-densities, and Razin-Tsytovich suppression only requires thermal plasma in the vicinity of the emission region. The existence of a characteristic rest-frame low-frequency cutoff therefore may rely on there being a common FRB progenitor, rather than a common progenitor environment.

\subsection{Predictions and future measurements}

The presence of a characteristic low-frequency spectral cutoff for FRBs has specific predictions that enable this scenario to be tested. These tests can also lead to a measurement of the cutoff frequency, $\nu_{\rm peak}$, given some knowledge or assumptions about the form of the spectrum, and the FRB luminosity function. Such tests are important, because the observations discussed above can have other explanations. For example, even if the ASKAP spectral-index measurement \citep{msb+18} is in error, an alternative explanation for the GBT and MWA non-detections is a characteristically flat FRB spectrum, such as that inferred for the repeating FRB\,121102 between 1--8\,GHz \citep{gsp+18}. 

Once a large FRB sample with redshift measurements becomes available, the mean rest-frame spectrum can be estimated by stacking redshift-corrected FRB spectra. If the difference between the Parkes and ASKAP FRB samples can be explained by a negative $K$-correction boosting the Parkes FRB rate at higher redshifts, the FRBs typically detected around 1.4\,GHz need to be observed below or near the rest-frame spectral peaks. This would also be consistent with the GBT/MWA non-detections at low frequencies. The required large FRB samples will be provided by localization observations with ASKAP, and by the Deep Synoptic Array; we note that bandpass calibration of these observations even at the off-boresight beam locations will be critical. These measurements will also enable the FRB luminosity function to be estimated, which needs to have a form such that fainter events are typically more cosmologically distant. 

Prior to the assembly of a large sample of FRBs with redshifts, insight into the existence of a low-frequency spectral cutoff will be provided by an analysis of the detection rate at different frequencies. A unique prediction of our hypothesis is the existence of a specific observing frequency with a maximum FRB detection rate for a fixed fluence threshold. As shown in Figure~\ref{fig:2}, this will typically occur at a frequency somewhat below the rest-frame $\nu_{\rm peak}$, when the negative $K$-correction maximizes the detection volume. If the negative observed spectral index estimated by \citet{msb+18} is a true indicator of the rest-frame spectral indices of the ASKAP sample, the GBT and MWA constraints would imply that the FRB detection rate may peak around the CHIME frequency band. The rate of CHIME detections in sections of their large fractional bandwidth will be particularly revealing. On the other hand, if ASKAP is instead observing FRBs below or near their rest-frame $\nu_{\rm peak}$, the FRB rate may peak at frequencies above 1.4\,GHz. 

We have also shown that a characteristic low-frequency spectral cutoff for FRBs will result in substantively different redshift and fluence distributions for different observing frequencies. Observations at low frequencies, below the frequency with the peak FRB rate, will preferentially detect more distant FRBs than observations at higher frequencies (Figure~\ref{fig:3}). This can result in low-frequency observations revealing FRB samples with relatively flat fluence distributions close to their detection thresholds (Figure~\ref{fig:4}). Observations at higher frequencies will preferentially detect events at lower redshifts, revealing steeper or Euclidean-space fluence distributions. The magnitudes of these effects will depend on the steepness of the rest-frame FRB spectrum both below and above $\nu_{\rm peak}$.

\section{Conclusions} \label{sec:5}

We have examined the possibility of a low-frequency cutoff in the characteristic rest-frame spectrum of FRBs. We conclude the following:
\begin{enumerate}

\item A selection of effects can result in the absorption or suppression of FRB emission at low frequencies, regardless of the specific emission mechanism (Figure~\ref{fig:1}). The extreme brightness temperatures of FRBs means that induced Compton scattering will occur for all reasonable thermal-plasma progenitor environments. Stimulated Raman-scattering interactions with Langmuir waves and Razin-Tsytovich suppression will also occur for dense plasma (e.g., $n_{e}\sim10^{6}$\,cm$^{-3}$ for $T_{e}\sim10^{6}$\,K). Free-free absorption may be relevant in particularly dense, cold environments. Examples of low-frequency spectral cutoffs exist in practice among the Galactic pulsar population. The potentially complex effects of the $K$-correction, beyond the power-law models previously considered \citep{vrh+16,me18b}, must therefore be incorporated into predictions of cosmological-FRB population models \citep[cf.][]{fl17}. 

\item A characteristic low-frequency spectral cutoff for FRBs will, under a variety of assumptions, manifest in the FRB rate being maximized for a particular observing frequency (Figure~\ref{fig:2}). Relative to higher frequency observations, surveys below the maximal-rate frequency will preferentially detect higher-redshift events (Figure~\ref{fig:3}), and will result in samples with sharply broken fluence distributions (Figure~\ref{fig:4}). High-frequency observations will be more likely to detect nearby events. These results are more pronounced for sharper spectral cutoffs, with observations below the maximal-rate frequency yielding more faint high-redshift events, and fewer bright nearby events. 

\item We suggest that the differences between the Parkes and ASKAP FRB samples, together with the non-detections of FRBs at low frequencies, can be explained by the suppression of low-frequency FRB emission even if the population does not evolve with cosmic time. A difficulty with this scenario is the steeply negative spectral index that may be characteristic of ASKAP FRBs \citep{msb+18}. Our hypothesis will be tested by measurements of the FRB rate and fluence distribution at multiple frequencies, in particular by CHIME. If FRB DMs form a good proxy for cosmological redshift, the FRB DM distributions at different frequencies will also be revealing. Direct measurements of the characteristic FRB rest-frame spectrum with FRBs localized to host galaxies with redshift measurements will provide a further test.

\end{enumerate}

\acknowledgments

VR thanks R. Blandford for raising the possibility of stimulated Raman scattering in FRB plasma environments. This work was supported in part by grants from the Breakthrough Prize Foundation and the Black Hole Initiative through the John Templeton Foundation.

\end{document}